\documentclass[12pt]{article}
\usepackage{epsfig}
\usepackage{color}
\usepackage{lineno}

\begin{document}

\begin{center}

\vspace*{1.0cm}

{\large \bf{New limits on double-beta decay of $^{190}$Pt and
$^{198}$Pt}}

\vskip 0.5cm

{\bf F.A.~Danevich$^{a,}$\footnote{Corresponding author. {\it
E-mail address:} danevich@kinr.kiev.ua (F.A.~Danevich).},
M.~Hult$^{b}$, A.~Junghans$^{c}$, D.V.~Kasperovych$^{a}$,
B.N.~Kropivyansky$^{a}$, G.~Lutter$^{d}$, G.~Marissens$^{b}$,
O.G.~Polischuk$^{a}$, M.V.~Romaniuk$^{a}$, H.~Stroh$^{b}$,
S.~Tessalina$^{e}$, V.I.~Tretyak$^{a}$, B.~Ware$^{e}$}

\vskip 0.3cm

$^{a}${\it Institute for Nuclear Research of NASU, 03028 Kyiv,
Ukraine}

$^{b}${\it European Commission, Joint Research Centre, Retieseweg
111, 2440 Geel, Belgium}

$^{c}${\it Helmholtz-Zentrum Dresden-Rossendorf, Bautzner
Landstrasse 400 01328 Dresden, Germany}

$^{d}${\it Department of Environmental Engineering, Technical
University of Denmark, DTU Ris\o~Campus, 4000, Roskilde, Denmark}

$^{e}${\it John de Laeter Centre for Isotope Research, GPO Box U
1987, Curtin University, Bentley, WA, Australia}

\end{center}

%\linenumbers

\vskip 0.5cm

\begin{abstract}
A search for double-beta decay of $^{190}$Pt and $^{198}$Pt with
emission of $\gamma$-ray quanta was realized at the HADES
underground laboratory with a 148 g platinum sample measured by
two ultralow-background HPGe detectors over 8946 h. The isotopic
composition of the platinum sample has been measured with high
precision using inductively coupled plasma mass spectrometry. New
lower limits for the half-lives of $^{190}$Pt relative to
different channels and modes of the decays were set on the level
of $\lim T_{1/2}\sim 10^{14}-10^{16}$ yr. A possible  exact
resonant $0\nu KN$ transition to the 1,2 1326.9 keV level of
$^{190}$Os is limited for the first time as $T_{1/2} \geq 2.5
\times10^{16}$ yr. A new lower limit on the double-beta decay of
$^{198}$Pt to the first excited level of $^{198}$Hg was set as
$T_{1/2} \geq 3.2\times10^{19}$ yr, one order of magnitude higher
than the limit obtained in the previous experiment.
\end{abstract}

\vskip 0.4cm

\section{Introduction}

Double-beta decay was considered for the first time by
Goeppert-Mayer in 1935 \cite{Goeppert-Mayer:1935}. The neutrino
accompanied mode of the decay with emission of electrons
(two-neutrino double-beta decay, $2\nu2\beta^-$) is allowed in the
Standard Model of particles and interactions (SM) and is already
observed in eleven nuclei with measured half-lives in the range
$T_{1/2}\sim (10^{19}-10^{24})$ yr \cite{Barabash:2020}. Another
possibility could be $2\beta^-$ decay without neutrino emission,
neutrinoless double-beta decay ($0\nu2\beta^-$). However, this
process violates the lepton number conservation and is only
possible if neutrinos are massive Majorana particles
\cite{Schechter:1982}. Thus, the $0\nu2\beta^-$ decay is the most
sensitive test of the lepton number conservation law and one of
the most promising tools to study properties of the neutrino and
the weak interaction. In general, the neutrinoless process can be
mediated by many effects beyond the SM and is considered as one of
the most powerful probes of the SM
\cite{Deppisch:2012,Bilenky:2015,DellOro:2016,Dolinski:2019}.
Despite the attempts, the $0\nu2\beta^-$ decay is still not
observed: the most sensitive experiments give upper half-life
limits in the range $\lim T_{1/2}\sim (10^{24}-10^{26})$ yr. This
allows to bound the effective Majorana neutrino mass at the level
of $\lim \langle m_{\nu}\rangle\sim (0.1-0.5)$ eV (see, e.g., the
reviews
\cite{DellOro:2016,Dolinski:2019,Giuliani:2012,Cremonesi:2014} and
the recent original works \cite{Gando:2016,
Anton:2019,Alvis:2019,Azzolini:2019,Agostini:2020,Adams:2020,Armengaud:2021}).

Other channels of double-beta decay are ``double-beta plus'' decay
processes: double-electron capture ($2\varepsilon$), electron
capture with positron emission ($\varepsilon\beta^+$), and
double-positron decay ($2\beta^+$). Although allowed in the SM,
the two-neutrino mode of the $2\varepsilon$ process (theoretically
the fastest decay channel due to the biggest values of the phase
space factor) has yet to be indisputably discovered. There are
indications of $2\nu2K$ decay with $T_{1/2}\sim 10^{21}$ yr in
$^{130}$Ba \cite{Meshik:2001,Pujol:2009,Meshik:2017} and $^{78}$Kr
\cite{Gavrilyuk:2013,Ratkevich:2017}. However, the indications for
$^{130}$Ba $2\nu2K$ decay was obtained only in geochemical
experiments by detection of an anomaly in the traces of xenon
isotopic concentration in barite minerals. The results of the
experiment \cite{Gavrilyuk:2013,Ratkevich:2017} with a
proportional chamber need to be confirmed. Recently an observation
of the $2\nu2K$ decay of $^{124}$Xe with $T_{1/2}=(1.8 \pm 0.5)
\times 10^{22}$ yr was claimed by the XENON collaboration by using
the XENON1T dark-matter detector \cite{XENON:2019}. As for the
neutrinoless double-beta plus decay processes, their mechanisms
are the same as for the decay with electrons emission. Moreover,
there are some additional arguments to develop experimental
methods to search for the $0\nu2\varepsilon$,
$0\nu\varepsilon\beta^+$ and $0\nu2\beta^+$ decays, taking into
account the potential to investigate the possible contribution of
the right-handed currents to the $0\nu2\beta^-$ decay rate if
observed \cite{Hirsch:1994}, and an interesting possibility of a
resonant $0\nu2\varepsilon$ process (see \cite{Blaum:2020} and
references therein).

\begin{figure}
\centering
 \mbox{\epsfig{figure=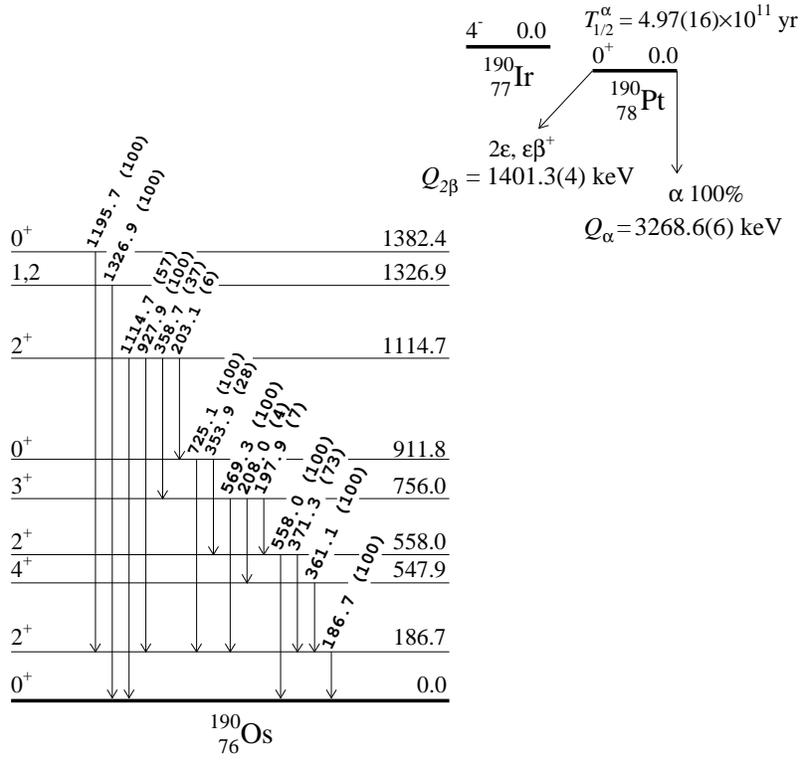,height=10.0cm}}
\caption{A simplified decay scheme of $^{190}$Pt \cite{NDS190}.
The energies of the excited levels and of the emitted $\gamma$
quanta are in keV (the relative intensities of the $\gamma$ quanta
are given in parentheses). $Q_{2\beta}$ and  $Q_{\alpha}$ are the
double-beta decay and alpha decay energies of $^{190}$Pt,
respectively.}
 \label{fig:190Pt}
\end{figure}

Two platinum isotopes are potentially unstable relative to the
double-beta decay: $^{190}$Pt with the decay energy
$Q_{2\beta}=1401.3(4)$ keV \cite{Wang:2021} and the isotopic
abundance $\delta=0.012(2)\%$ \cite{Meija:2016}, and $^{198}$Pt
with the decay energy $Q_{2\beta}=1050.3(21)$ keV \cite{Wang:2021}
and the isotopic abundance $\delta=7.356(130)\%$
\cite{Meija:2016}. A simplified decay scheme of $^{190}$Pt is
shown in Fig. \ref{fig:190Pt}.
\noindent The nuclide remains one of few nuclides with a
possibility of exact resonant $0\nu2\varepsilon$ decay. The
resonant character of the process is expected in the $0\nu KN$
transition to the 1,2 1326.9(10) keV excited level of the daughter
\cite{Blaum:2020}. It should be stressed that the decay channel
was not considered in the previous work \cite{Belli:2011} due to a
rather poor accuracy of the $Q_{2\beta}$ value at the time of the
experiment (in 2011 the recommended $Q_{2\beta}$ was 1383(6) keV
\cite{Audi:2003}). A status of the decay as a potential resonance
can be characterized by the resonance parameter $R_f$:

\begin{equation}
 R_f = \Gamma_f/(\Delta^2 + \Gamma_f^2/4),
 \label{eq:res-par}
\end{equation}

\noindent where $\Gamma_f = \Gamma_1 + \Gamma_2$ is the
de-excitation width of the electron shell of the daughter nuclide.
The degeneracy parameter $\Delta$ is equal to
$Q_{2\beta}-E_{exc}-E_{b1}-E_{b2}$, where $E_{exc}$ is the energy
of the daughter nucleus excited level, $E_{bi}$ are the binding
energies of the captured electrons on the atomic shells of the
daughter atom. Taking into account the presently recommended
$Q_{2\beta}$ value \cite{Wang:2021} the resonance parameter for
$^{190}$Pt (normalized on the value for the $0\nu2\varepsilon$
decay $^{54}$Fe $\rightarrow$ $^{54}$Cr) can reach one of the
biggest values $\approx7.0\times10^{8}$ among the possible
resonant transitions \cite{Blaum:2020}\footnote{Unfortunately, the
1326.9-keV level characteristics are known with a rather limited
accuracy: the uncertainty of the level energy is $\pm1.0$ keV,
spin is 1 or 2, while parity of the level is unknown.}. The method
of ultralow-background HPGe $\gamma$-ray spectrometry was applied
in the present study to search for different channels and modes of
$2\varepsilon$ and $\varepsilon\beta^+$ processes in $^{190}$Pt,
including the possible resonant transitions in the nuclide.

\begin{figure}
\centering
 \mbox{\epsfig{figure=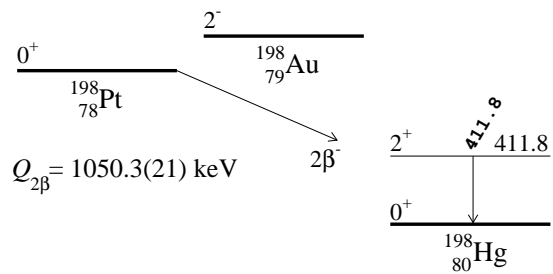,height=3.5cm}}
\caption{A simplified decay scheme of $^{198}$Pt \cite{NDS198}.
The energies of the excited level and of the emitted $\gamma$
quantum are in keV. $Q_{2\beta}$ is the double-beta decay energy
of $^{198}$Pt.}
 \label{fig:198Pt-scheme}
\end{figure}

A simplified decay scheme of another possible $2\beta$ unstable
isotope of platinum, $^{198}$Pt, is presented in Fig.
\ref{fig:198Pt-scheme}. The $2\beta^-$ transition of $^{198}$Pt to
the first $2^+$ 411.8 keV excited level of $^{198}$Hg is expected
to be accompanied by emission of $\gamma$-ray quanta with energy
411.8 keV that can be detected by $\gamma$-spectrometry methods
too.

In the next Section we describe the high-purity platinum sample,
precise measurements of its isotopic composition, and the ultralow
background HPGe detector system used in the experiment. The data
analysis and obtained limits on the $2\varepsilon$ and
$\varepsilon\beta^+$ processes in $^{190}$Pt and the $2\beta^-$
decay of $^{198}$Pt to the first excited level of the daughter are
presented in Section \ref{sec:res-dis}. A summary of the
experiment is given in the Conclusions section.

\section{Experiment}
\label{sec:experiment}

\subsection{Platinum sample, isotopic composition of the material}
\label{sec:abundance}

A disk-shaped sample of metallic platinum with a diameter of
25.04(1) mm, a thickness of 14.07(2) mm, and with a mass of
148.122(1) g was used in the experiment. The purity grade of the
platinum is 99.95\%\footnote{In May 2016 the Pt disk was used as a
neutron transmission target in the Institute of Radiation Physics
of the Helmholtz-Zentrum (Dresden). It was $\approx1$ m away from
the electron beam (30 MeV) and the neutron producing radiator. The
target was irradiated by a neutron flux of about $200-400$
neutrons per second for about one week. The neutron spectrum
starts around $10-100$ keV and ends at around 10 MeV with a
maximum around $1-2$ MeV.}. The representative isotopic abundance
of $^{190}$Pt in normal terrestrial materials has a rather big
uncertainty $\delta=0.012(2)\%$ \cite{Meija:2016}. Thus, special
mass-spectrometry measurements of the sample were realized.

The Pt isotopic measurements were acquired using a sector field ICP-MS ELEMENT XR (ThermoScientific) at the John de Laeter Centre for Isotope Research, Curtin University. Measurements of masses $^{190}$Pt,$^{192}$Pt, $^{194}$Pt, $^{195}$Pt, $^{196}$Pt, and $^{198}$Pt were performed in low resolution mode using electrostatic scanning (e-scan, i.e., peak jumping) from a set magnet mass at $^{190}$Pt. Due to the drastically different abundances of Pt isotopes, the ELEMENT XR’s triple detection mode is advantageous for such analyses as isotopes such as $^{190}$Pt and $^{192}$Pt can be measured in a pulse-counting detection mode, while the remaining masses can be analysed in analogue mode all within the same analytical session. Prior to the analysis of each sample, a blank solution of 2\% HNO$_{3}$ was measured to correct for background. A summary of the platinum isotopic composition, as well as numbers of nuclei of the isotopes in the sample are presented in Table \ref{tab:abundance}.

\clearpage

\begin{table}
\caption{Isotopic composition ($\delta$) of the platinum sample
measured in the present work and the numbers of nuclei of each
isotope in the sample calculated by using the measured isotopic
concentrations. The combined standard uncertainties of the
isotopic abundances are given with a coverage factor $k=2$
(approximately 95\% level of confidence). The representative
isotopic abundances from \cite{Meija:2016} are given too.}
 \centering
%\label{tab:abundance}
\begin{tabular}{cccc}
\hline\noalign{\smallskip}

 Isotope            & \multicolumn{2}{c}{$\delta$ (\%)} & Number of nuclei\\
                    & IUPAC \cite{Meija:2016}   & this work     & in the sample   \\
\noalign{\smallskip}\hline\noalign{\smallskip}

$^{190}$Pt    & 0.012(2)                        & 0.0127(1)     & $5.81(5)\times10^{19}$ \\
~                   &                           & ~             & ~                       \\
$^{192}$Pt    & 0.782(24)                       & 0.7759(16)    & $3.548(7)\times10^{21}$ \\
~                   &                           & ~             & ~                       \\
$^{194}$Pt    & 32.864(410)                     & 32.6511(522)  & $1.4929(24)\times10^{23}$ \\
~                   &                           & ~             & ~                       \\
$^{195}$Pt    & 33.775(240)                     & 33.6884(526)  & $1.5403(24)\times10^{23}$ \\
~                   &                           & ~             & ~                       \\
$^{196}$Pt    & 25.211(340)                     & 25.5376(419)  & $1.1677(19)\times10^{23}$ \\
~                   &                           & ~             & ~                       \\                   &               & ~            & ~                       \\
$^{198}$Pt    & 7.356(130)                      & 7.3343(115)   & $3.353(5)\times10^{22}$ \\

\noalign{\smallskip}\hline
\end{tabular}
\label{tab:abundance}
\end{table}

\subsection{Ultralow-level gamma-ray spectrometry measurements}
 \label{sec:low-bg-msr}

\begin{figure}[ht]
\centering
\includegraphics[width=0.4\textwidth]{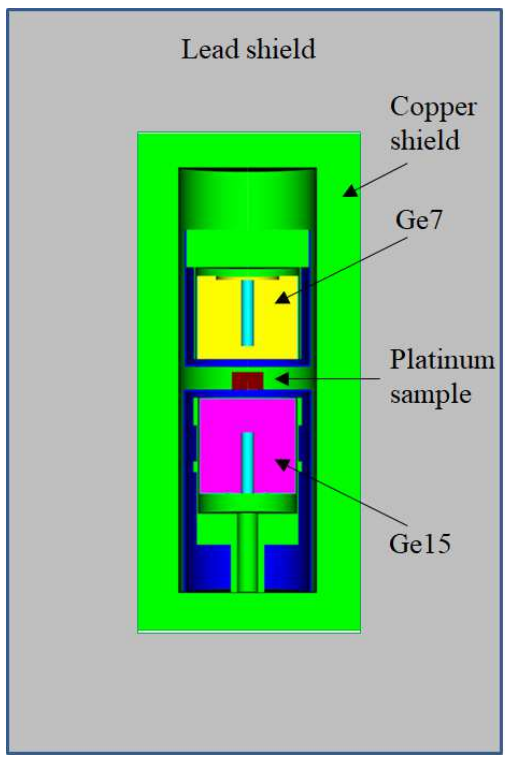}
\caption{Schematic of the experimental ultralow-background set-up
with the HPGe detectors and the platinum sample.}
 \label{fig:set-up-sketch}
\end{figure}

The platinum sample was measured in an ultralow-background
HPGe-detector system located 225 m underground in the laboratory
HADES (Belgium) \cite{Andreotti:2010,Hult:2020}. The detector
system consists of two p-type Extended Range HPGe-detectors facing each other (see a
schematic of the set-up in Fig. \ref{fig:set-up-sketch}). Both the detectors were manufactured by Canberra semiconductor (Olen, Belgium). The
detectors were shielded by 35 mm electrolytic copper (innermost)
then 40 mm ultralow-level lead and on the outside 145 mm lead. The
main characteristics of the HPGe detectors are presented in Table
\ref{tab:detectors}, more details of the detector system can be
found in \cite{Wieslander:2009}.

\clearpage

\begin{table}[htb]
\caption{Properties of the HPGe-detectors used in the present
experiment. FWHM denotes the full width at half of maximum of
$\gamma$-ray peak measured with a $^{60}$Co gamma-ray source.}
\begin{center}
\begin{tabular}{|l|c|c|}
 \hline
 ~                                      & Ge7           & Ge15 \\
 \hline

 Energy resolution (FWHM) at 1333 keV   & 2.2 keV       & 1.8 keV \\

 Relative efficiency                    & 90\%          & 85\% \\

 Crystal mass                           & 1778 g        & 1840 g \\

 Endcap / Window material               & HPAl / HPAl   & HPAl / HPAl \\

 Dead layer (front)                     & 0.3 $\mu$m   & 0.3 $\mu$m  \\
 \hline
 
 % Manufacturer							& \multicolumn{2}{l|}{Canberra semiconductor, Olen, Belgium}  \\
%\cline{2-3}
  \hline

 \multicolumn{3}{l}{HPAl = High Purity Aluminum} \\

\end{tabular}
\label{tab:detectors}
\end{center}
\end{table}

%\clearpage

The data with the Pt sample were collected starting from February
6th, 2018 over a period of 805 days in 6 runs covering a live time
of 8946 h (373 d). The measurements of background without sample
were carried out for 674 h (28 d). The energy calibration and
stability check was performed using reference point sources
containing $^{60}$Co, $^{137}$Cs and $^{241}$Am.

\section{Results and discussion}
\label{sec:res-dis}

\subsection{Radionuclides detected in the platinum sample}
\label{sec:rad-cont}

\begin{figure}[ht]
\centering
\includegraphics[width=0.7\textwidth]{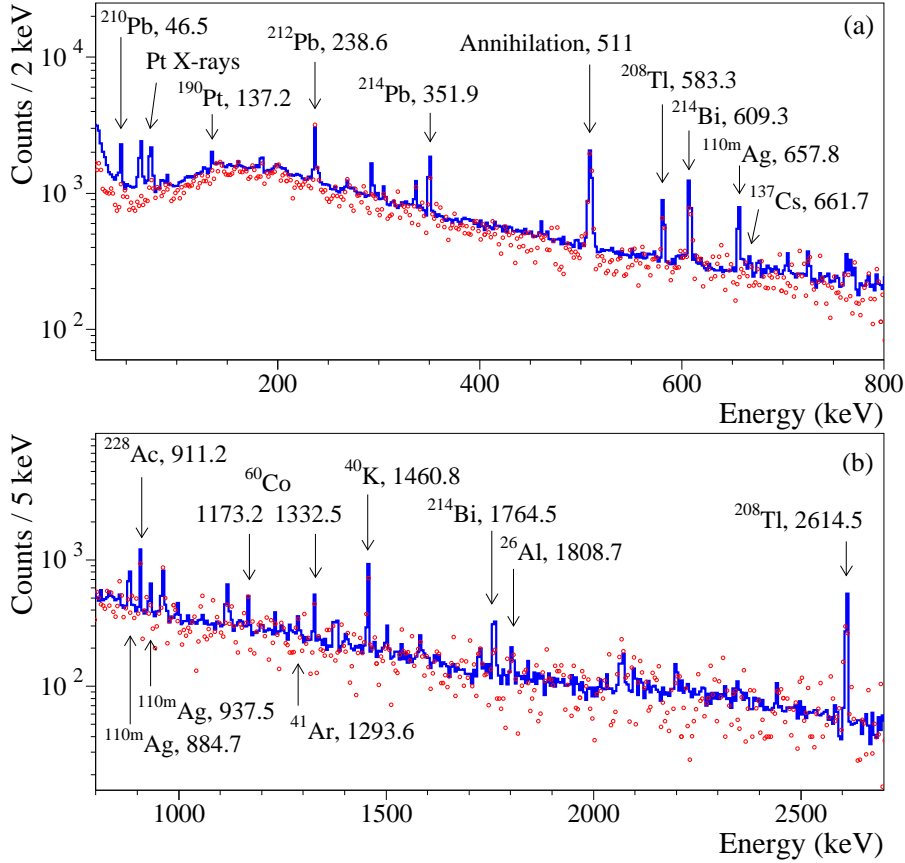}
\caption{Energy spectra in the energy intervals $20-800$ keV (a)
and $800-2700$ keV (b) measured with the platinum sample over 8946
h (solid histogram) and without sample over 674 h (normalized to
8946 h, dots) by the ultralow-background HPGe-detector system. The
energy of the $\gamma$ peaks is in keV.}
 \label{fig:BG}
\end{figure}

The sum energy spectra measured by the Ge7 and Ge15 detectors with
the Pt sample and without sample (background) are shown in Fig.
\ref{fig:BG}. The majority of the peaks can be assigned to
$^{40}$K and nuclides of the $^{232}$Th, $^{235}$U and $^{238}$U
decay families. There are also peaks of $^{22}$Na, $^{26}$Al,
$^{54}$Mn, $^{60}$Co, $^{137}$Cs and $^{110m}$Ag. The traces of
$^{22}$Na, $^{26}$Al and $^{110m}$Ag that were detected are
presumingly reminiscence from neutron activation of minor
impurities of the Pt-sample during the neutron experiment in
Dresden and during air-transport. Radioactive $^{22}$Na and
$^{26}$Al can be also cosmogenically generated in the aluminum
details of the HPGe detectors \cite{Majorovits:2011}. $^{54}$Mn
and $^{60}$Co are typical cosmogenic radionuclides that can be
produced in Pt, copper and some other materials of the set-up.
Presence of $^{137}$Cs can be a result of the set-up (sample)
pollution after the Chernobyl or (and) Fukushima Daiichi nuclear
disasters. In the Pt data there is a clear $\gamma$ peak with
energy 137.2 keV due to the $\alpha$ decay of $^{190}$Pt to the
137.2 keV excited level of $^{186}$Os
\cite{Belli:2011a}\footnote{The results of the $\alpha$ decay
investigation will be presented in a separate report.}. Also peaks
due to neutron-gamma reactions on the materials of the set-up were
observed: in particular, a 139.7-keV peak of $^{75m}$Ge produced
by neutron-gamma reaction on $^{74}$Ge, a 198.4-keV peak from
$^{70}$Ge(n,$\gamma$)$^{71m}$Ge reaction, a 202.6-keV peak from
$^{115}$In(n,$\gamma$)$^{116}$In reaction. There is also a peak of
$^{41}$Ar with energy 1293.6 keV due to operation of the BR-1
nuclear reactor of the Belgian nuclear research centre. The peak
is present on few specific days (approximately 15 days in total)
during the measurements when air blows from the reactor towards
the inlet of the ventilation for the HADES laboratory. We have
decided to do not exclude the data with the $^{41}$Ar peak taking
into account a rather mild effect of the radioactivity in the
energy intervals of interest.

\nopagebreak

The energy dependence of the energy resolution in the sum energy
spectrum measured with the Pt sample by the detectors Ge7 and Ge15
was determined for the low energy region ($65-352$ keV) by using
intense X-ray and $\gamma$-ray peaks with energies 65.1 keV and
66.8 keV ($K_{\alpha2}$ and $K_{\alpha1}$ X-ray of Pt), 137.2 keV
($^{190}$Pt), 238.6 keV ($^{212}$Pb), 295.2 keV and 351.9 keV
($^{214}$Pb) as:

\begin{equation}
 \mathrm{FWHM(keV)}=1.08(12)+0.020(8)\sqrt{ E_{\gamma}},
 \label{eq:fwhm1}
\end{equation}

\noindent where $E_\gamma$ is in keV. For the energies above 352
keV we use an approximation obtained by analysis of the $\gamma$
peaks 238.6 keV ($^{212}$Pb), 295.2 keV and 351.9 keV
($^{214}$Pb), 583.2 keV ($^{208}$Tl), 609.3 keV ($^{214}$Bi),
1173.2 keV and 1332.5 keV ($^{60}$Co) and 1460.8 keV ($^{40}$K):

\begin{equation}
 \mathrm{FWHM(keV)}=0.55(8)+0.049(4)\sqrt{ E_{\gamma}}.
 \label{eq:fwhm2}
\end{equation}

The specific activity\footnote{In this paper ``specific activity'' is the activity per unit mass of the sample and not to the mass of a certain isotope.} of all detected radionuclides was calculated
using the following formula:

\begin{equation}
A = (S_{sample}/t_{sample}-S_{bg}/t_{bg})/(\epsilon \cdot \eta
\cdot m),
 \label{eq:rad-cont}
\end{equation}

\noindent where $S_{sample}$ ($S_{bg}$) is the area of a peak in
the sample (background) spectrum; $t_{sample}$ ($t_{bg}$) is the
time of the sample (background) measurement; $\epsilon$ is the
$\gamma$-ray emission intensity of the corresponding transition;
$\eta$ is the full energy peak efficiency; $m$ is the sample mass.
The detection efficiencies were calculated with the EGSnrc
simulation package \cite{Kawrakow:2017,Lutter:2018}, the events
were generated homogeneously in the Pt sample. The Monte Carlo
models of the two detectors have being validated through several
participations in proficiency tests and an additional validation
measurement was done using a $^{57}$Co point $\gamma$ source. The
standard deviation of the relative difference between the
simulations and the experimental data is $3.3\%$ for $\gamma$-ray
peaks with energies 122.1 keV and 136.5 keV for the detectors Ge7
and Ge15. The estimated specific activities of radioactive
impurities in the platinum are presented in Table
\ref{tab:rad-cnt}.

\nopagebreak
\begin{table}[ht]
\caption{The specific activities of the detected radionuclides. The
upper limits are given at 90\% confidence level (C.L.), the
reported uncertainties are the combined standard uncertainties calculated by summing the systematic and statistical uncertainties in quadrature.
The reference date is the start of the measurement (February 6th,
2018).}
\begin{center}
\begin{tabular}{|l|l|l|}

 \hline
  Chain     & Nuclide       &  Specific activity (mBq/kg) \\
 \hline
 ~          & $^{22}$Na     & $\leq 0.5$ \\
 ~          & $^{26}$Al     & $\leq 0.6$ \\
 ~          & $^{40}$K      & $\leq 13$ \\
 ~          & $^{54}$Mn     & $\leq 0.9$ \\
 ~          & $^{60}$Co     & $\leq 0.8$ \\
 ~          & $^{137}$Cs    & $\leq 0.4$ \\
 ~          & $^{110m}$Ag   & $9.2\pm0.4$ \\
 \hline
 $^{232}$Th & $^{228}$Ra   & $\leq 5.7$ \\
 ~          & $^{228}$Th   & $10.8\pm1.3$ \\
 \hline
 $^{235}$U  & $^{235}$U    & $\leq 16$  \\
 ~          & $^{231}$Pa   & $\leq 81$  \\
 \hline
 $^{238}$U  & $^{234}$Th  & $247\pm 134$ \\
  ~         & $^{234m}$Pa  & $\leq 89$ \\
  ~         & $^{226}$Ra   & $6.8\pm0.9$  \\
  ~         & $^{210}$Pb   & $2600\pm 570$   \\
  \hline
\end{tabular}
\label{tab:rad-cnt}
\end{center}
\end{table}

\clearpage

\subsection{Limits on the double-beta decay processes in $^{190}$Pt
} \label{sec:190Pt-limits}

No peculiarity was observed in the experimental energy spectra
that could be ascribed to the $2\beta$ decay processes in
$^{190}$Pt or $^{198}$Pt. Thus, we set limits on different modes
and channels of the decays by using the following formula:

\begin{equation}
\lim T_{1/2} = N \cdot \eta \cdot t \cdot \ln 2 / \lim S,
 \label{eq:limT12}
\end{equation}

\noindent where $N$ is the number of nuclei of interest in the
sample (see Table \ref{tab:abundance}), $\eta$ is the detection
efficiency for the $\gamma$-ray (X-ray) quanta searched for, $t$
is the measuring time, and $\lim S$ is the number of events of the
effect which can be excluded at a given C.L. In the present work
all the $\lim S$ values and the half-life limits are given with
90\% C.L. The detection efficiencies of the detector system to the
$\gamma$ (X-ray) quanta expected in different modes and channels
of the double-beta processes in $^{190}$Pt and $^{198}$Pt were
simulated with the EGSnrc simulation package
\cite{Kawrakow:2017,Lutter:2018}, the decay events were generated
by the DECAY0 events generator \cite{DECAY0}.

\begin{figure}[ht]
\centering
\includegraphics[width=0.7\textwidth]{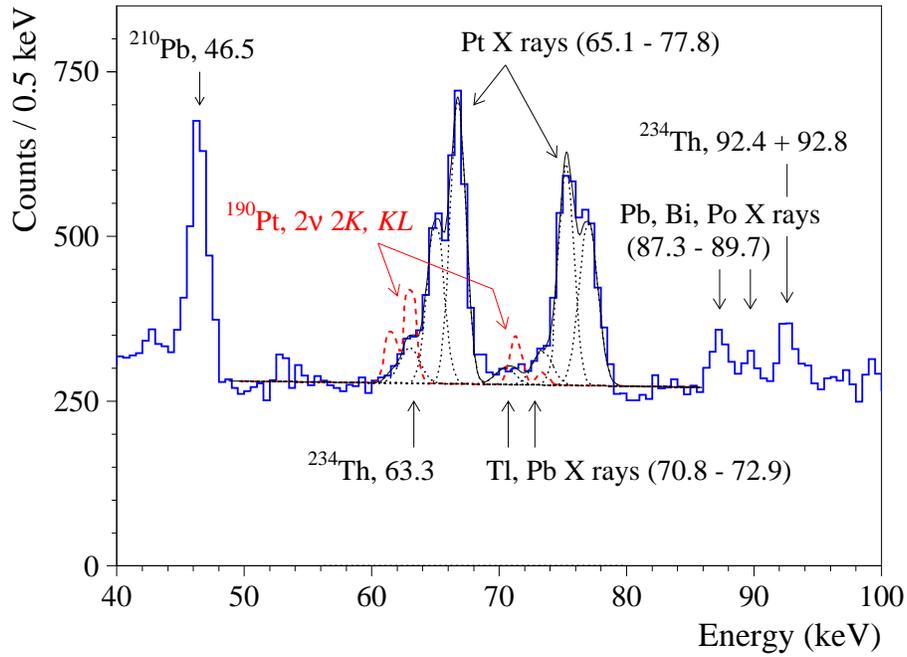}
\caption{The energy spectrum collected with the Pt sample by the
HPGe detector system in the energy region where $K$ X-ray quanta
are expected for the $2\nu 2K$ and $2\nu KL$ decays of $^{190}$Pt.
The fit of the data by the background model is shown by solid
line, the peaks included in the model of the background are shown
by dotted lines, while the excluded effect is presented by dashed
line (the excluded distribution is multiplied by a factor 2 to
improve visibility). Energy of $\gamma$ and X-ray quanta are in
keV.}
 \label{fig:2n2K}
\end{figure}

A cascade of X-rays and Auger electrons due to deexcitation of the Os
electron shell with individual energies in the energy interval
$\approx (61.5-73.4)$ keV is expected in the $2\nu2K$ and $2\nu
KL$ capture in $^{190}$Pt. However, the energies of the $L$ X-rays
are very low, $(7.8 - 12.5)$ keV, and therefore heavily attenuated
by the sample and also by the aluminium-windows of the
HPGe-detectors. The Auger electrons avoid detection for the same
reason. Thus, the response of the detector system to the $2\nu2K$
and $2\nu KL$ decays of $^{190}$Pt was built assuming the
following energies and intensities of X-rays from the $K$ shell of the Os atom (only the X-rays with the intensities higher than 0.5\%
were considered): 61.5 keV ($K_{\alpha 2}, 27.5\%$), 63.0 keV
($K_{\alpha 1}, 47.3\%$), 71.1 keV ($K_{\beta 3}, 5.45\%$), 71.4
keV ($K_{\beta 1}, 10.50\%$), 73.4 keV ($K_{\beta 2}, 3.69\%$)
\cite{TOI}. The detection efficiencies to the X-ray quanta were
simulated by the EGSnrc code. To estimate $\lim S$ values for the
$2\nu2K$ and $2\nu KL$ decays, the energy spectrum taken with the
Pt sample was fitted in the energy interval $49-86$ keV by the
$2\nu2K$ ($2\nu KL$) distribution\footnote{The $2\nu2K$ and $2\nu
KL$ distributions are almost similar in the region of fit.} and a
sum of several Gaussian functions to describe the 63.3-keV
$\gamma$-ray peak of $^{234}$Th, X-ray peaks of Pt, Tl, Pb, Bi and
Po present in the spectrum, plus a straight line to describe the
continuous distribution. The result of the fit is shown in Fig.
\ref{fig:2n2K}. Despite an empirical approach to build the background model, the quality of the fit is good: $\chi^2$/n.d.f.$ =
39.1/50=0.78$ (where n.d.f. is a number of degrees of freedom).
The fit returned an area of the $2\nu2K$ ($2\nu KL$) distribution
$S=164\pm125$ counts that indicates no evidence of the effect
searched for. An excluded effect can be estimated by the simple Gaussian approach as $\lim S=369$ counts at 90\% C.L. Taking into
account the Monte-Carlo simulated detection efficiency for the
whole $2\nu2K$ ($2\nu KL$) effect $\eta=0.0092$ ($\eta=0.0047$),
the following half-life limits were obtained for the $2\nu2K$ and
$2\nu KL$ capture in $^{190}$Pt:

\begin{center}
 $T_{1/2}^{2\nu 2K}~(^{190}$Pt, g.s.$~\rightarrow~$g.s.)~$\geq~1.0\times 10^{15}$ yr,
\end{center}

\begin{center}
 $T_{1/2}^{2\nu KL}~(^{190}$Pt, g.s.$~\rightarrow~$g.s.)~$\geq~5.2\times10^{14}$
 yr.
\end{center}

The half-life limits, the energies of the X-ray quanta
($E_{\gamma}$), which were used to set the $T_{1/2}$ limits, the
detection efficiencies ($\eta$) and values of $\lim S$ are
presented in Table \ref{table:limits}\footnote{Only statistical errors originating from the data fluctuations were taken into account in the present study in the estimations of the half-life limits, systematic effects have not been included in the half-life limits evaluations.}, where also results of the
previous experiment \cite{Belli:2011} are given for comparison.
The limit for the $2\nu2K$ decay half-life slightly exceeds the
one reported in the previous work \cite{Belli:2011}, while the
limit for the $2\nu KL$ capture is obtained for the first time.

\nopagebreak

\begin{table}[ht]
\caption{The half-life limits on $2\varepsilon$ and
$\varepsilon\beta^+$ processes in $^{190}$Pt and $2\beta^-$ decay
of $^{198}$Pt to the first excited level of $^{198}$Hg. The
energies of the X-ray or $\gamma$ quanta ($E_{\gamma}$), which
were used to set the $T_{1/2}$ limits, are listed with their
corresponding detection efficiencies ($\eta$) and values of $\lim
S$. The results of previous experiment \cite{Belli:2011} are given
for comparison. Values of $\lim S$ and $T_{1/2}$ are given at 90\%
C.L. All the empirical fits from which the signal upper limits are derived describe the data well, with the goodness-of-fit values within $\chi^2$/n.d.f.$=(0.39-0.84)$.}
\begin{center}
\begin{tabular}{lllllll}

\hline
 Transition                         & Level of          & $E_{\gamma}$      & $\eta$    & $\lim S$  & \multicolumn{2}{c}{Experimental limit} \\
                                    & daughter          & (keV)             & ~         & (counts)  & \multicolumn{2}{c}{$T_{1/2}$ (yr)} \\
 ~                                  & nucleus (keV)     & ~                 & ~         & ~         & Present work            & \cite{Belli:2011} \\
 \hline
 $^{190}$Pt$~\to$ $^{190}$Os        & ~                 & ~                 & ~         & ~         & ~                       & ~  \\
 $2\nu 2K$                          & g.s.              & $61.5-73.4$       & 0.0092    & 369       & $\geq1.0\times10^{15}$  & $\geq8.4\times10^{14}$ \\
 $2\nu KL$                          & g.s.              & $61.5-73.4$       & 0.0047    & 369       & $\geq5.2\times10^{14}$  & ~ \\
 $2\nu 2\varepsilon$                & $2^+$ 186.7       & 186.7             & 0.0093    & 39        & $\geq9.8\times10^{15}$  & $\geq8.8\times10^{14}$$~^*$ \\
 $2\nu 2\varepsilon$                & $2^+$ 558.0       & 558.0             & 0.0285    & 50        & $\geq2.3\times10^{16}$  & $\geq5.6\times10^{15}$ \\
 $2\nu 2\varepsilon$                & $0^+$ 911.8       & 725.1             & 0.0393    & 31        & $\geq5.2\times10^{16}$  & $\geq4.5\times10^{15}$ \\
 $2\nu 2\varepsilon$                & $2^+$ 1114.7      & 1114.7            & 0.0145    & 31        & $\geq1.9\times10^{16}$  & $\geq1.0\times10^{16}$ \\
 $2\nu KN$                          & 1,2 1326.9        & 1326.9            & 0.0491    & 81        & $\geq2.5\times10^{16}$  & ~ \\

 $0\nu 2K$                          & g.s.              & $1253.2-1254.0$   & 0.0500    & 44       & $\geq4.7\times10^{16}$  & $\geq5.7\times10^{15}$ \\
 $0\nu KL$                          & g.s.              & $1314.1-1317.0$   & 0.0493    & 44       & $\geq4.6\times10^{16}$  & $\geq1.7\times10^{16}$ \\
 $0\nu 2L$                          & g.s.              & $1375.0-1380.0$   & 0.0484    & 62       & $\geq3.2\times10^{16}$  & $\geq3.1\times10^{16}$ \\

 $0\nu 2\varepsilon$                & $2^+$ 186.7       & 186.7             & 0.0072    & 39        & $\geq7.6\times10^{15}$  & $\geq6.9\times10^{14}$ \\
 $0\nu 2\varepsilon$                & $2^+$ 558.0       & 558.0             & 0.0234    & 50        & $\geq1.9\times10^{16}$  & $\geq4.5\times10^{15}$ \\
 $0\nu 2\varepsilon$                & $0^+$ 911.8       & 725.1             & 0.0361    & 31        & $\geq4.8\times10^{16}$  & $\geq3.6\times10^{15}$ \\
 $0\nu 2\varepsilon$                & $2^+$ 1114.7      & 1114.7            & 0.0117    & 31        & $\geq1.6\times10^{16}$  & $\geq9.8\times10^{15}$ \\
 Res. $0\nu KN$                     & 1,2 1326.9        & 1326.9            & 0.0488    & 81        & $\geq2.5\times10^{16}$  & ~ \\
 Res. $0\nu LM$                     &(0,1,2)$^+$ 1382.4 & 1195.7            & 0.0497    & 42        & $\geq4.9\times10^{16}$  & $\geq2.9\times10^{16}$ \\

 $2\nu\varepsilon\beta^+$           & g.s.              & 511               & 0.0953    & 134       & $\geq2.9\times10^{16}$  & $\geq9.2\times10^{15}$ \\
 $0\nu\varepsilon\beta^+$           & g.s.              & 511               & 0.0942    & 134       & $\geq2.9\times10^{16}$  & $\geq9.0\times10^{15}$ \\
 $2\nu\varepsilon\beta^+$           & $2^+$ 186.7       & 511               & 0.0926    & 134       & $\geq2.8\times10^{16}$  & $\geq8.4\times10^{15}$ \\
 $0\nu\varepsilon\beta^+$           & $2^+$ 186.7       & 511               & 0.0922    & 134       & $\geq2.8\times10^{16}$  & $\geq8.4\times10^{15}$ \\

 ~                                  &~                  &~                  & ~         & ~         & ~        & ~                 \\
 $^{198}$Pt$~\to$ $^{198}$Hg &~                  &~                  & ~         & ~         & ~        & ~                       \\
 $2\beta^{-}$ ($2\nu+0\nu$)         & $2^+$ 411.8       & 411.8             & 0.0414   & 31        &  $\geq3.2\times10^{19}$ & $\geq3.5\times10^{18}$  \\

 \hline

 \multicolumn{7}{l}{$^*$ The limit in \cite{Belli:2011} was set for the $2\nu2K$ transition to the 186.7 keV level of $^{190}$Os.} \\
% \multicolumn{7}{l}{$^{**}$ The limit was obtained for another energy of $\gamma$-ray quanta $E_{\gamma}$} \\
 \end{tabular}
 \label{table:limits}
 \end{center}
 \end{table}

 \clearpage

\begin{figure}[ht]
\centering
\includegraphics[width=0.6\textwidth]{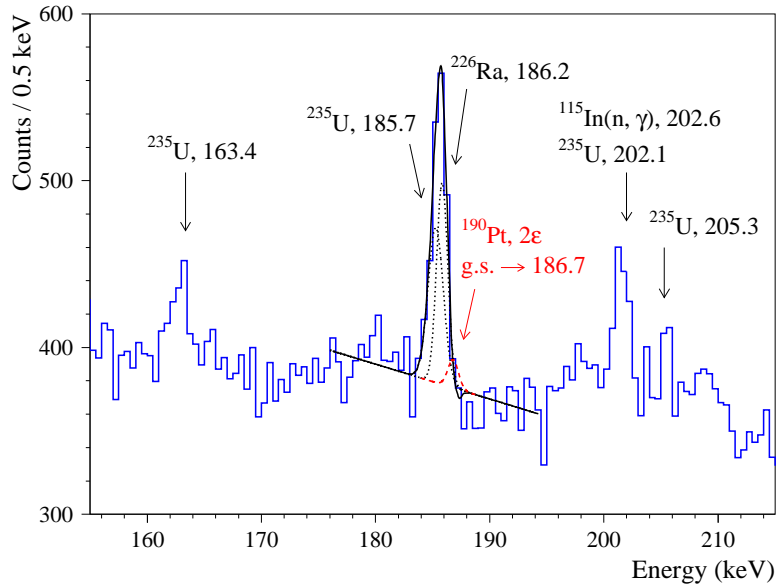}
\caption{The energy spectrum gathered with the Pt sample in the
vicinity of $\gamma$ peak 186.7 keV expected in the $2\varepsilon$
decay of $^{190}$Pt to the $2^+$ 186.7 keV excited level of
$^{190}$Os. The fit of the data is shown by solid line, while the
excluded peak is presented by dashed line. The dotted lines show
the peaks of $^{235}$U and $^{226}$Ra included in the background
model. The energies of the background and expected $\gamma$ peaks
are in keV.}
 \label{fig:186.7keV}
\end{figure}

The $2\nu$ double-electron capture in $^{190}$Pt can undergo to several
excited levels of $^{190}$Os (see Fig. \ref{fig:190Pt}). In this
case, in addition to X-rays cascade, $\gamma$-peaks at the energy
of the excited level (in the transition to the first $2^+$ 186.7
keV excited level) and also at the energies of the transitions
between the initial and final levels of the daughter are expected.
To estimate $\lim S$ value for a possible 186.7-keV peak area, the
experimental spectrum was fitted in the energy interval $176-194$
keV by a model consisting of a straight line, peaks of $^{235}$U
and $^{226}$Ra with energies 185.7 keV and 186.2 keV,
respectively, and a Gaussian peak at 186.7 keV with the width
fixed according to the formula (\ref{eq:fwhm1}) to describe the
effect searched for. The fit returns a 186.7-keV peak area
$S=-28(39)$ counts, that corresponds to $\lim S=39$ counts
following the recommendations \cite{Feldman:1998}. A part of the
energy spectrum gathered with the Pt sample in the vicinity of
$\gamma$ peak 186.7 keV expected in the $2\varepsilon$ decay of
$^{190}$Pt to the $2^+$ 186.7 keV excited level of $^{190}$Os is
presented in Fig. \ref{fig:186.7keV}. The half-life limits for the
transitions to higher excited levels of $^{190}$Os were obtained
in a similar way. The limits are presented in Table
\ref{table:limits}.

\nopagebreak

\begin{figure}[ht]
\centering
\includegraphics[width=0.6\textwidth]{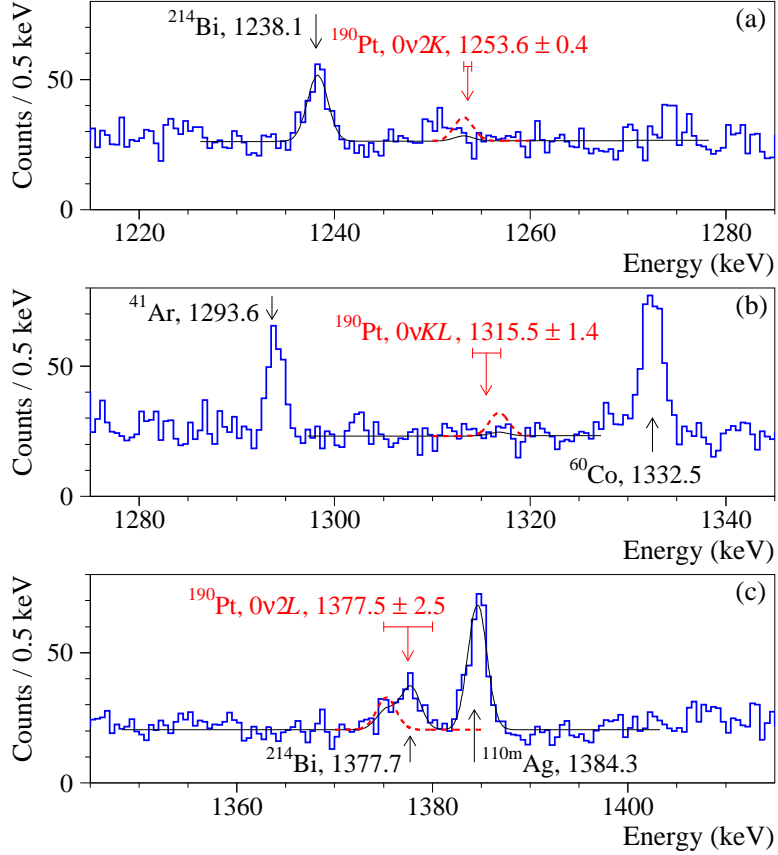}
\caption{The energy spectrum gathered with the Pt sample in the
energy intervals where the $\gamma$ peaks from the $0\nu2K$ (a),
$0\nu KL$, and $0\nu 2L$ (c) captures in $^{190}$Pt to the
ground state of $^{190}$Os are expected. The fits of the data are
shown by solid lines (see text for details of the background models), while the excluded peaks are presented by
dashed lines. The horizontal lines (above the arrows labelling the
energy of the peaks searched for) show the energy intervals
corresponding to the uncertainties of the expected energies of the
peaks. The energy of the $\gamma$ peaks are in keV.}
 \label{fig:0n2e}
\end{figure}

In the $0\nu$ double-electron capture in $^{190}$Pt the energy
excess is assumed to be emitted by bremsstrahlung $\gamma$-ray
quanta with an energy
$E_{\gamma}=Q_{2\beta}-E_{b1}-E_{b2}-E_{exc}$. To estimate values
of $\lim S$ for the $0\nu$ double-electron captures from $K$ and
$L$ shells, the experimental spectrum was fitted in the regions of
the expected peaks with energies (for the g.s. to g.s. transition)
$1253.6 \pm 0.4$ keV, $1315.5 \pm 1.4$ keV and $1377.5 \pm 2.5$
keV for the $0\nu2K$, $0\nu KL$, and $0\nu 2L$ captures,
respectively. The variations of the expected peaks energy are due
to the $Q_{2\beta}$ value uncertainty and different binding
energies of the $L$ atomic shells. Thus, the energy of a peak
searched for was free parameter of the fits within the variations. A straight line was taken to describe the continuous background. The
background model included background $\gamma$ peaks  present in the
energy intervals of the fits: 1238.1 keV of $^{214}$Bi (to estimate  a $\lim S$ for the $0\nu2K$ decay), and 1377.7 keV of $^{214}$Bi plus 1384.3 keV of $^{110m}$Ag (in the case of the $0\nu2L$ capture). The results of the fits are shown in
Fig. \ref{fig:0n2e}. The biggest peak areas were taken to derive
the $\lim S$ values for the peaks searched for (see Table
\ref{table:limits}).

\nopagebreak

Limits on the neutrinoless $2\varepsilon$ transitions to the
excited levels of the daughter were set by using the $\lim S$
values obtained to estimate the half-life limits for the
two-neutrino processes. However, the detection efficiencies for
the $0\nu$ modes are slightly smaller due to the energy transfer
to the bremsstrahlung $\gamma$-ray quanta (absent in the $2\nu$
decay mode, where the energy is carried out by neutrinos).

The $0\nu KN$ transition in $^{190}$Pt to the 1,2 1326.9 keV
excited level of $^{190}$Os is of special interest: the decay can
be an exactly resonant, up to six orders of magnitude faster, if
the degeneracy parameter $\Delta =
Q_{2\beta}-E_{exc}-E_{b1}-E_{b2} \sim 10$ eV. However, with the
uncertainties of the decay energy ($\pm 0.4$ keV \cite{Wang:2021})
and of the 1326.9-keV level energy ($\pm 1.0$ keV \cite{NDS190})
the degeneracy parameter lies in the interval from $-0.13$ keV to
$+0.26$ keV with a combined uncertainty $\pm1.1$ keV, that is too
big to make a clear conclusion about resonance enhancement of the
transition. Nevertheless, development of experimental techniques
to study the decay at, as much as possible, high sensitivity level
is an important task. The energy spectrum gathered with the Pt
sample was fitted in the energy interval $1310-1357$ keV to
estimate a half-life limit on the decay. The background was
described by a straight line and by a peak at 1332.5 keV
($\gamma$-peak of $^{60}$Co). The position of the peak searched
for was bounded within $\pm1$ keV, while the peak width was fixed
as FWHM~$=2.33$ keV according to the formula (\ref{eq:fwhm2}). The
fit (with $\chi^2/$n.d.f.$=41.4/88=0.47$) returned a peak area
$S=29\pm32$ counts\footnote{If no bounds were set to the peak
parameters, the energy of the peak was $1328.1(4)$ keV, the peak
width FWHM~$=2.28(40)$ keV, and the peak area
 $S=41(22)$ counts.}. Considering the peculiarity as a
statistical fluctuation we have obtained $\lim S=81$ counts. Taking
into account the detection efficiency for $\gamma$-ray quanta with
energy 1326.9 keV ($\eta=0.0488$), the resonant $0\nu KN$
transition in $^{190}$Pt to the 1,2 1326.9 keV excited level of
$^{190}$Os is limited as:

\begin{center}
 $T_{1/2}^{0\nu KN}~(^{190}$Pt, g.s.$~\rightarrow~1326.9$ keV$)~\geq~2.5\times10^{16}$ yr.
\end{center}

\begin{figure}[ht]
\centering
\includegraphics[width=0.6\textwidth]{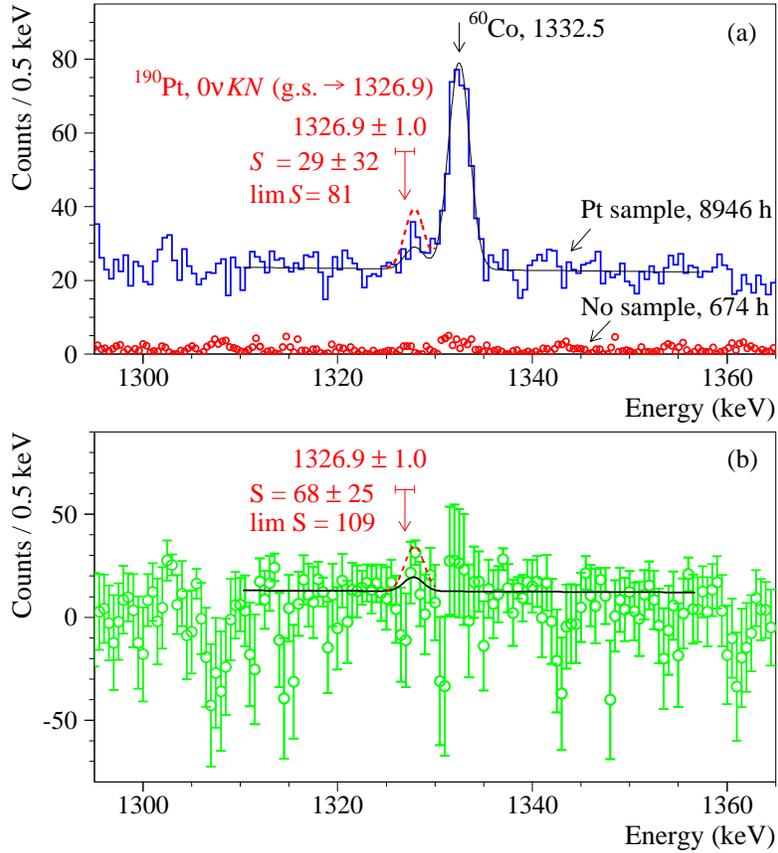}
\caption{The energy spectra gathered with the Pt sample (with no sample) by the detector system where the $\gamma$ peak from possible resonant
$0\nu KN$ transition to the 1,2 1326.9 keV level of $^{190}$Os is
expected (a). The background-subtracted spectrum is shown in panel (b). The fits of the data are shown by solid lines (see text for details of the background models), while the excluded
peaks are presented by dashed lines.}
 \label{fig:res1}
\end{figure}

The energy spectrum in the region of interest, the result of the
fit and excluded peak for the possible resonant $0\nu KN$
transition in $^{190}$Pt to the 1,2 1326.9-keV excited level of
$^{190}$Os is shown in Fig. \ref{fig:res1} (a). 

The bound on the transition to the 1326.9 keV level of $^{190}$Os is limited by $^{60}$Co background, that can be attributed entirely to the setup contamination (see Table \ref{tab:rad-cnt}) and not to the Pt sample. One can try to estimate a limit on the transition by analysis of the no-sample spectra subtracted from the data taken with the Pt sample, eliminating in such a way the $^{60}$Co background. A fit of the background-subtracted spectrum by a straight line (to describe the continuous background) plus a Gaussian peak with the bounded position and fixed width (the effect searched for) is shown in panel (b) of Fig. \ref{fig:res1}\footnote{A fit of the data with no bounds on the peak position and width returned the peak area $S=34(17)$ counts and the peak position 1327.8(5) keV, however, with an abnormally small peak width FWHM~$=0.4(3)$ keV.}. However, statistical fluctuations in the background-subtracted spectrum are rather big due to a comparatively short time of the background measurement. As a result, the limit from the background-subtracted spectrum ($\lim S=109$ counts) is worse than the one from the sample-only spectrum. Nevertheless, the both approaches give comparable results providing a useful cross-check of the analysis.

\nopagebreak

 \clearpage

\begin{figure}[ht]
	\centering
	\includegraphics[width=0.6\textwidth]{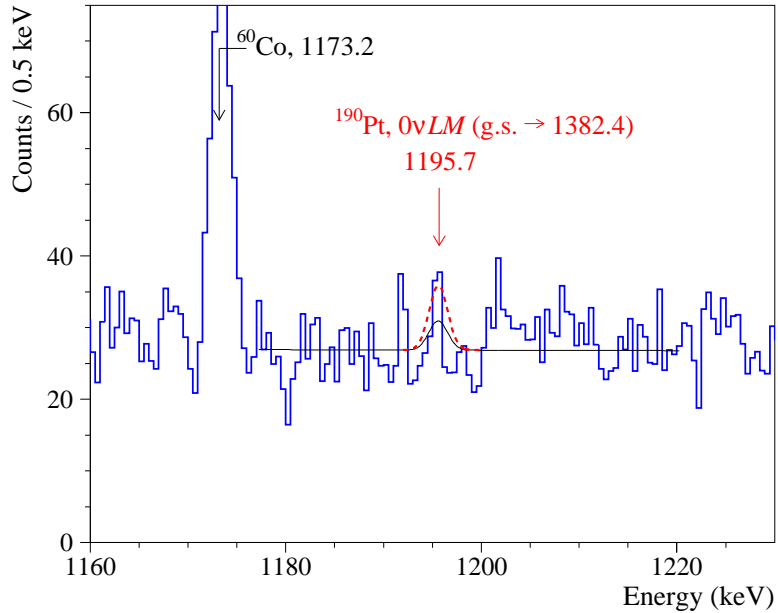}
	\caption{The energy spectrum in the vicinity of $\gamma$ peak
		1195.7 keV expected in the near resonant $0\nu LM$ transition of
		$^{190}$Pt to the $(0,1,2)^+$ 1382.4 keV level of $^{190}$Os (b).
		The fit of the data is shown by solid line, while the excluded
		peaks are presented by dashed line.}
	\label{fig:res2}
\end{figure}

The $0\nu LM$ transition of $^{190}$Pt to the $(0,1,2)^+$
1382.4(2) keV level of $^{190}$Os can be characterized as ``near''
resonant one since the degeneracy parameter in this case is rather
big $\Delta = (2.9-6.1)\pm0.6$ keV. The estimation of half-life
limit for the decay was performed by analysis of the experimental
data in the vicinity of $\gamma$ peak with energy 1195.7(2) keV
emitted in the de-excitation of the 1382.4 keV level. The result
of the data fit in the energy interval $1177-1220$ keV by a simple
model constructed from a straight line and a Gaussian peak at
1195.7(2) keV with a fixed width is presented in Fig.
\ref{fig:res2}. 

\nopagebreak

\begin{figure}[htb]
\begin{center}
 \mbox{\epsfig{figure=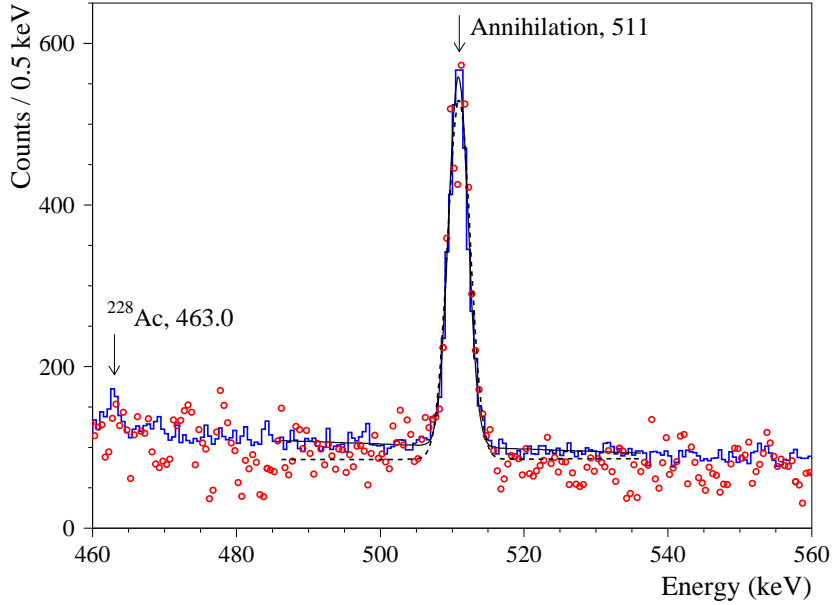,height=8.0cm}}
\caption{The energy spectrum measured with the Pt sample for 8946
h in the vicinity of the 511 keV annihilation peak. The background
data (measured over 674 h, normalized on 8946 h) are shown by
dots. The fit of the data obtained with the Pt sample is shown by
solid line, the fit of the background spectrum is presented by
dashed line.}
 \label{fig:511}
 \end{center}
 \end{figure}

The annihilation peaks at 511 keV were analyzed in the data with
(without) the Pt sample to estimate half-life limits for the
electron capture with positron emission in $^{190}$Pt to the
energetically allowed transitions to the ground state and to the
excited 186.7 keV level of the daughter. A fit of the data with
(without) the Pt sample gives the 511-keV peak area $S=3143 \pm
71$ counts for measurement time 8946 h ($S=266 \pm 18$ counts  for
674 h) that leads to the difference: $-388\pm249$ counts that
results in $\lim S=134$ counts. The energy spectrum taken with the
Pt sample and the background spectrum in the vicinity of the 511
keV annihilation peak with the results of fits are shown in Fig.
\ref{fig:511}. The detection efficiencies are slightly different
for the decays to the ground state and to the first 186.7 keV
excited level of the daughter, as well as for the $2\nu$ and
$0\nu$ modes of the decays. The obtained half-life limits are
presented in Table \ref{table:limits}.

\nopagebreak

The obtained limits on the possible double-beta processes in
$^{190}$Pt are higher up to an order of magnitude than the limits
set in the previous study \cite{Belli:2011}. Moreover, the
expected peaks positions for the $0\nu2\varepsilon$ decays were
far from the actual ones, taking into account a rather different
$Q_{2\beta}$ value and much bigger uncertainty at the time when
the experiment \cite{Belli:2011} was carried out. For this reason,
the half-life limits for the $0\nu 2K$, $0\nu KL$ and $0\nu 2L$
captures can be considered as obtained for the first time.

However, the sensitivity of the present study is very far from
theoretical estimations of the $^{190}$Pt decay probability. For
instance, the calculations of the $^{190}$Pt half-life relative to
the resonant $0\nu2\varepsilon$ transitions to the 1326.2 keV
[1382.4 keV] level of $^{190}$Os are in the interval $T_{1/2}\sim
(3.3\times10^{26}-1.6\times10^{30})$ yr [$T_{1/2}\sim
(1.0\times10^{30}-6.5\times10^{30})$ yr] (assuming the effective
Majorana neutrino mass $\langle m_{\nu}\rangle=0.1$ eV and the
weak axial-vector coupling constant $g_A=1.27$) (see Table IV in
\cite{Blaum:2020}). Nevertheless, achievement of the sensitivity
on the level of $T_{1/2}\sim (10^{26}-10^{27})$ yr looks realistic
taking into account the great progress in the ultralow-background
HPGe \cite{Agostini:2020}  and low-temperature bolometers
\cite{Azzolini:2019,Adams:2020,Armengaud:2021,Alenkov:2019}
detection techniques, the experimental approaches that look the
most promising for possible large-scale searches for the resonant
$0\nu2\varepsilon$ decays.

\subsection{Limit on the $2\beta^-$ decay of $^{198}$Pt to the first
excited level of $^{198}$Hg} \label{sec:198Pt-limit}

\begin{figure}[!htbp]
 \begin{center}
 \mbox{\epsfig{figure=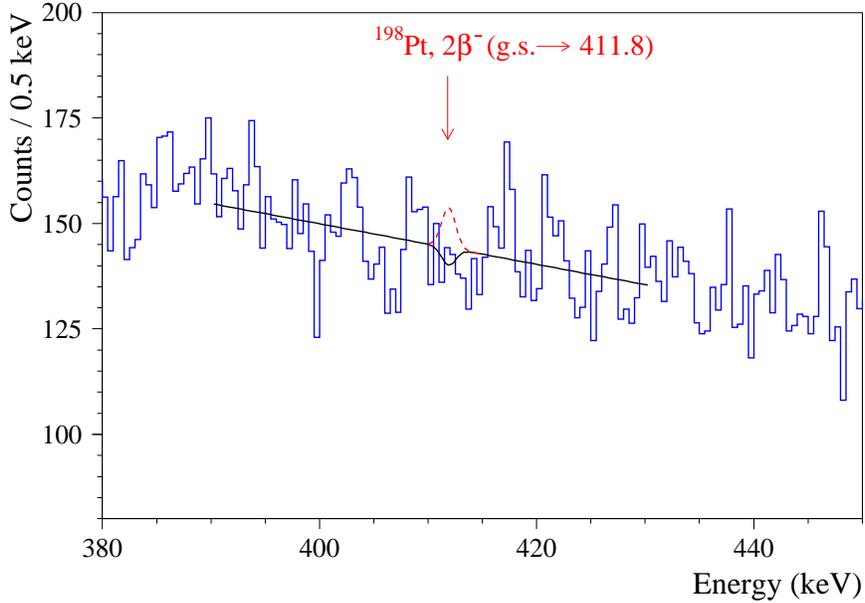,height=8.0cm}}
\caption{Part of the energy spectrum measured with the Pt sample
where a peak with energy 411.8 keV after the $2\beta^-$ decay of
$^{198}$Pt to the first $2^+$ 411.8 keV excited level of
$^{198}$Hg is expected. The fit of the data is presented by solid
line, the excluded $\gamma$ peak is shown by dashed line.}
 \label{fig:198Pt}
 \end{center}
 \end{figure}

A limit on the $2\beta^-$ transition of $^{198}$Pt to the $2^+$
411.8 keV excited level of $^{198}$Hg was set by analysis of the
energy spectrum measured with the Pt sample. The spectrum was fit
in the energy interval $390-430$ keV by a straight line (to
describe the continuous background) and a Gaussian function at
411.8 keV with the peak width determined by the formula
(\ref{eq:fwhm2}) returning an effect area $S=-13\pm 26$ counts,
that corresponds to $\lim S=31$ counts (the energy spectrum in the
region of the fit is presented in Fig. \ref{fig:198Pt}). Taking
into account the detection efficiency ($\eta= 0.0414$ both for the
$2\nu$ and $0\nu$ modes of the decay), the following limit was
obtained:

\begin{center}
$T_{1/2}^{(2\nu+0\nu)2\beta^{-}}(^{190}$Pt,
g.s.$~\rightarrow~411.8$ keV$)\geq~3.2\times10^{19}$ yr.
\end{center}

The limit is one order of magnitude higher than that set in the
previous experiment \cite{Belli:2011}. Theoretical calculations
for the $2\nu2\beta^-$ decay of $^{198}$Pt to the ground state of
the daughter are in the interval $T_{1/2}\sim
5\times10^{21}-5\times10^{23}$ yr \cite{Sta90,Hir94}. Similar
estimations give semiempirical formulae: $T_{1/2}=
(3.0\pm1.5)\times10^{23}$ yr \cite{Pri10} and $T_{1/2}=
3.3\times10^{22}$ yr \cite{Ren14}. A probability of the
$2\nu2\beta^-$ decay to the $2^+$ 411.8 keV excited level is
expected to be several orders of magnitude lower due to a smaller
value of the phase space factor and the spin change. The
theoretical estimations of the $0\nu2\beta^-$ decay of $^{198}$Pt
to the ground state of $^{198}$Hg are much higher: $T_{1/2}\sim
3\times10^{26}-2\times10^{28}$ yr
\cite{Sta90,Bar12,Bar13,Iac15}\footnote{The calculations were
performed for the effective Majorana neutrino mass $\langle
m_{\nu}\rangle=0.1$ eV and $g_A\approx1.27$.}, while the
$0\nu2\beta^-$ transition to the $2^+$ 411.8 keV excited level of
$^{198}$Hg is expected to be further suppressed. Thus, the
sensitivity of the present experiment is rather far from the
theoretical predictions for the two-neutrino mode of the decay,
not to say for the neutrinoless process.

All the limits obtained in the present experiment are summarized
in Table \ref{table:limits} where also results of the previous study
 \cite{Belli:2011} are given for comparison. 
 
\section{Summary and conclusions}

A high-purity disk-shaped platinum sample with mass 148 g was
measured in an ultralow background HPGe-detector $\gamma$-ray
spectrometry system located 225 m underground at the HADES
laboratory over 8946 hours aiming at searching for double-beta
decay of $^{190}$Pt and $^{198}$Pt with emission of $\gamma$-ray
quanta. The isotopic composition of the platinum sample has been
measured with high precision using inductively coupled plasma mass
spectrometry. No effect was observed but lower limits on the
half-lives for the different channels and modes of the decays of
$^{190}$Pt were set on the level of $\lim T_{1/2}\sim
10^{14}-10^{16}$ yr. A possible exact resonant $0\nu KN$
transition to the 1,2 1326.9 keV level of $^{190}$Os is limited
for the first time as $ T_{1/2} \geq 2.5 \times10^{16}$ yr. A new
improved limit is set for the $2\beta^-$ transition of $^{198}$Pt
to the $2^+$ 411.8 keV excited level of $^{198}$Hg as $ T_{1/2}
\geq 3.2 \times 10^{19}$ yr. All the obtained limits exceed the
previously obtained values up to one order of magnitude mainly
thanks to a substantially bigger exposure ($55.2$ kg$\times$day in
the present work and $3.2$ kg$\times$day in \cite{Belli:2011}).
However, the sample geometry was not optimised since it was an
existing sample foreseen for another project. Thus, a further
improvement of the present experiment sensitivity can be achieved
by decrease of the sample thickness by production of thin disk
with a diameter comparable to the detectors size.

The sensitivity of the experiment to the $^{190}$Pt decays,
particularly to the potentially resonant transition to the 1326.9
keV level of $^{190}$Os, could be advanced by using platinum
enriched in the isotope $^{190}$Pt, increasing the exposure and
detection efficiency by utilization of thin samples and
multi-crystal system of HPGe $\gamma$-ray detectors or
low-temperature bolometers. However, such an experiment could be
considered after more accurate determination of the 1326.9-keV
level energy (presently known with a rather big uncertainty
$\pm1.0$ keV). Nevertheless, we realize that implementation of
such an experiment is practically a rather difficult task, first
of all due to the inaccessibility of methods for enrichment of
platinum isotopes in the hundreds kilograms scale requested for a
competitive experiment (in terms of the Majorana neutrino mass)
even in a case of an exact resonant transition.

\section{Acknowledgments}

This work received support from the EC-JRC open access scheme
EUFRAT under Horizon-2020, project number 2018-35375-4 (PLATOS).
O.G.~Polischuk and M.V.~Romaniuk were supported in part by the
project ``Investigation of double-beta decay, rare alpha and beta
decays'' of the program of the National Academy of Sciences of
Ukraine ``Laboratory of young scientists'' (the grant number
0120U101838). F.A.~Danevich, D.V.~Kasperovych and V.I.~Tretyak
were supported in part by the National Research Foundation of
Ukraine Grant No. 2020.02/0011.

\end{document}